\documentclass[superscriptaddress,reprint, prl]{revtex4-2}

\usepackage{newtxtext,newtxmath}

\usepackage{graphicx}

\usepackage{dcolumn}

\usepackage{siunitx}
\usepackage[letterpaper,margin=1in]{geometry}

\date{\today}

\makeatletter
\renewcommand{\fnum@figure}{\textbf{Figure \thefigure}}
\renewcommand{\fnum@table}{\textbf{Table \thetable}}
\makeatother

\usepackage{url}

\begin{document} 

\title{
	Primordial Metamaterials
}

\author{
	A. Ware${\dagger}$}
\affiliation{The Chandra Department of Electrical and Computer Engineering, The University of Texas at Austin, Austin, 78712, USA.}

\author{J. LaMountain$^{\dagger}$}
\affiliation{Department of Physics and Applied Physics, University of Massachusetts Lowell, Lowell, 01854, USA.}

\author{R.C. White}
\affiliation{The Chandra Department of Electrical and Computer Engineering,  The University of Texas at Austin, Austin, 78712, USA.}
\affiliation{Department of Electrical and Computer Engineering,  University of Illinois Urbana-Champaign, Champaign and Urbana, 61801 USA.}

\author{S.R. Bank}
\affiliation{The Chandra Department of Electrical and Computer Engineering, The University of Texas at Austin, Austin, 78712, USA.}

\author{E. Narimanov}

\affiliation{Department of Electrical and Computer Engineering, Purdue University, West Lafayette, 47907, USA.}

\author{D. Wasserman}
\affiliation{The Chandra Department of Electrical and Computer Engineering, The University of Texas at Austin, Austin, 78712, USA.}

\author{V.A. Podolskiy}
\affiliation{Department of Physics and Applied Physics, University of Massachusetts Lowell, Lowell, 01854, USA.}
\email{viktor\_podolskiy@uml.edu}


\date{\today}

\begin{abstract} 
The electromagnetic response of materials serves as the foundation for a broad range of vital applications, including but not limited to imaging,  sensing, as well as classical and quantum communications. Here we demonstrate, theoretically and experimentally, a fundamentally new regime of electromagnetic material response originating from inherent material nonlocality, leading to effective ``spooky action at a distance''. We show that by structuring materials on the scale of their inherent nonlocality, it becomes possible to reveal the ``primordial'' nonlocal response of the components and design materials with strong overall nonlocality, easily detectable at room temperatures and in realistic (lossy) materials. Designer primordial nonlocality offers a new dimension in controlling light-matter interactions.
\end{abstract}

\maketitle


\noindent
Nonlocality, the ``spooky action at a distance''\cite{EPR,pekarJETP} of materials' electromagnetic response, typically reveals itself in exotic, difficult to achieve, regimes, such as entangled quantum systems, highly controlled nanoscale structures, or ultra-low-temperature single crystal materials\cite{EPR,Hopfield63,kiselevJETP73,deAbajo,norlanderPRB,schatzPRL,schatzPNASplasmonic,plasmonicBook,MortensenPRB}. The electromagnetic response of any material is inherently nonlocal, defined at the fundamental level by the quantum dynamics of its charges. While this leads to the polarization at a given point inside a material being dependent on the electric field at a neighboring point, the corresponding spatial scale of this inherent nonlocality is usually small. As a result, the electromagnetic response of the majority of everyday materials is not affected by nonlocal corrections. Here we propose and experimentally demonstrate that by spatially structuring materials it becomes possible to bring this “primordial” nonlocal electromagnetism back to the macroscopic scale. We realize primordial metamaterials in an epitaxially-grown, all-semiconductor, room-temperature materials platform. Our research presents new opportunities for shaping the optical response of matter: we are able to design the distribution of electromagnetic fields within, and the propagation of light through, these composite media by leveragng a novel and heretofore difficult-to-realize mechanism for controlling light-matter interaction. 

Materials' response to light can be described in terms of the materials' polarization distribution, fundamentally tied to the dynamics of charges within the materials\cite{landauECM,AgranovichBook}. The finite mass of these charges yields a time-lag between the driving external electric field and the resulting polarization distribution. This time-lag results in the dispersion (frequency-dependence) of polarizability, the function that relates the polarization within the material and the external electric field (see Fig.1a). This behavior generally is local -- polarization $\vec{P}$ at a given location inside the material depends only on the electric field $\vec{E}$ at that same point. However, at the nanoscale, where charge dynamics are non-trivial, materials' responses in adjacent positions couple to each other. For example, in free-electron-dominated plasmonic materials, electron motion due to the incident electromagnetic field induces polarization at the electron's current location -- potentially shifted from the electron's location at the time of field application . 

\begin{figure} 
	\centering
	\includegraphics[width=0.45\textwidth]{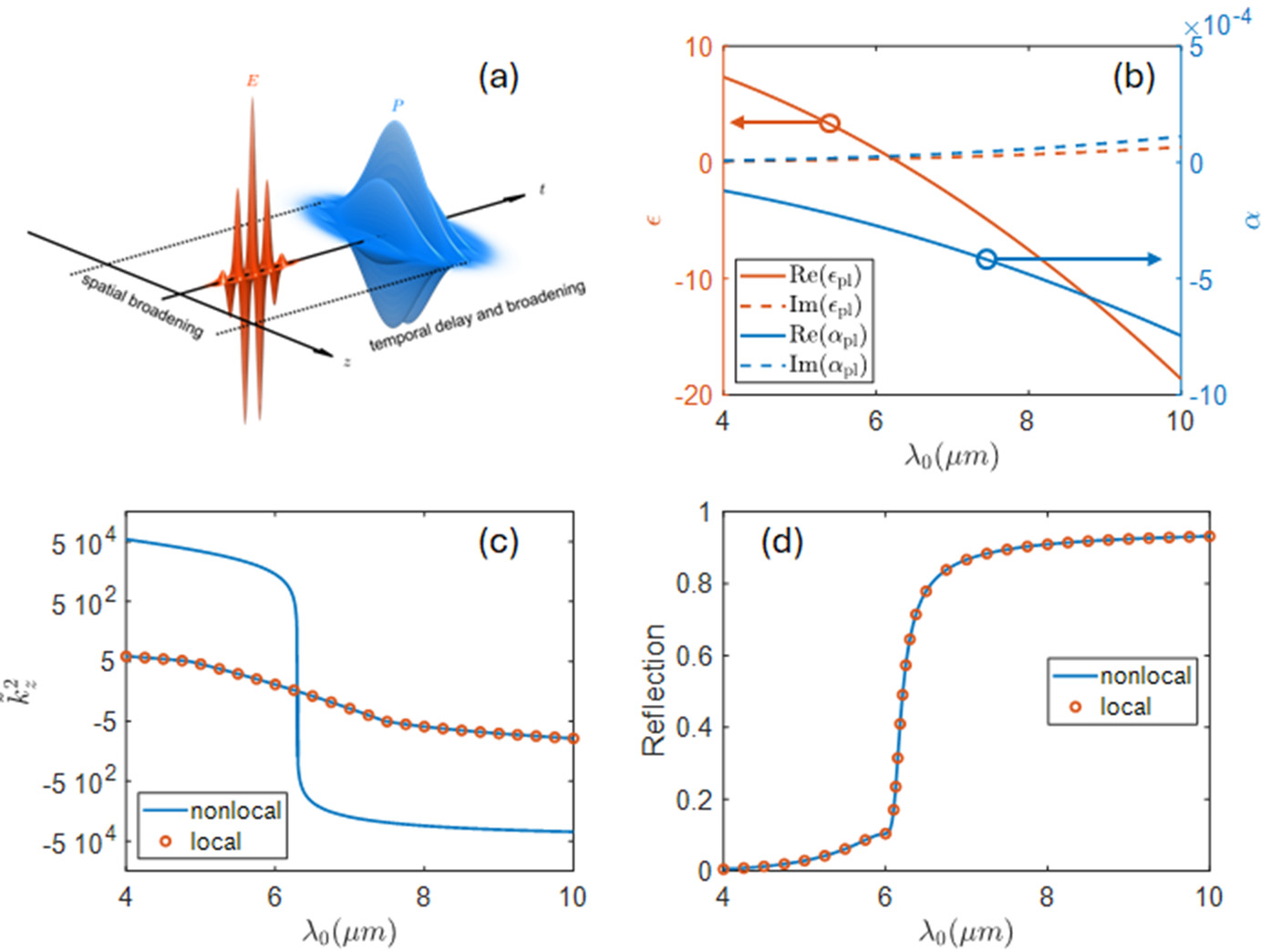} 

	\caption{\textbf{Local and nonlocal electromagnetism in bulk materials}
		(a) A material's response to an excitation pulse is often delayed in time and can affect the regions neighboring the excitation point (e.g. due to charge motion). The time-delay yields frequency dependence of polarizability (and equivalently, permittivity) while the spacial nonlocality yields dependence on wavenumber; (b) local and nonlocal parts of permittivity of a highly-doped plasmonic semiconductor, modeled via the Drude model; (c) dispersion of the plane waves propagating in a highly doped plasmonic medium, modeled with local (symbols) and nonlocal (lines) permittivity (note the nonlinear scale of the vertical axis); (d) reflectivity of a plasmonic layer (for light incident from vacuum at $60^o$), modeled via local (symbols) and nonlocal (line) models. }
	\label{figConfig} 
\end{figure}

Electromagnetic nonlocality, originally described in\cite{pekarJETP}, makes the polarizability dependent on wavenumber in addition to frequency.  Nonlocal behavior qualitatively changes materials' optical response by introducing additional electromagnetic waves. We illustrate this phenomenon using a highly doped  (plasmonic) semiconductor as a representative material. Fig.1(b) shows its dielectric permittivity while Fig. 1c illustrates the dispersion of the two waves in this material (characterized via dimensionalized propagation constant $\tilde{k}=k c/\omega$, with $\omega$ being angular frequency of light and $c$ being speed of light in vacuum). Note that the dispersion of one of the modes (primary wave) is nearly identical to the prediction of local electromagnetism, while the dispersion of the second (additional) wave can only be described by nonlocal calculations. At shorter wavelengths, both waves propagate ($\tilde{k}^2>0$) while at longer wavelengths, both waves exhibit evanescent decay ($\tilde{k}^2<0$). Notably, the magnitude of the propagation constant of the additional wave is significantly higher than that of its primary counterpart, virtually preventing the coupling of free-space radiation to nonlocality-driven additional modes. As a result, corrections due to nonlocality in the majority of real-life situations are vanishingly small.  As an example, Fig.\ref{figConfig}(d) illustrates reflection from the doped semiconductor, calculated using both local and nonlocal polarizability descriptions (see supplementary information, SI, for details of analytical and numerical derivations).  The local and nonlocal far-field optical responses of this representative system are essentially identical. 

The excitation of additional waves has been observed in a few exotic situations -  such as single-crystal phononic materials operating in deep cryogenic environments at epsilon-near-zero frequencies\cite{kiselevJETP73,Hopfield63}. While nonlocal electromagnetism can be mimicked in metamaterials and metasurfaces\cite{podolskiyNonlocalPRL,silveirinhaPRApplied,podolskiyNonlocalAPL,kivsharNonlocalEMT,kivsharNonlocalFishnet,robertsDIT,aluNonlocalMetasurface,yuNonlocalMetasurface,capassoNonlocalLaser,capassoNonlocalMetasurface}, the response of the underlying media in these situations remains local. Until now, additional waves have not been utilized for shaping nanoscale electromagnetic environments. In this work, we consider composite materials comprising multiple nonlocal components and present an approach to bring the underlying nonlocal response of materials to the forefront of their electromagnetic response.

\subsection*{Emergence of primordial waves}

The fundamental limitation to accessing and leveraging the nonlocality-related additional modes in any optical material lies in these modes' weak coupling to free-space radiation. Homogeneous materials exhibit measurable nonlocality only when their permittivity is vanishingly small so that propagation constants of nonlocal waves are small enough to interact with incoming light\cite{AgranovichBook}. Alternatively, composite structures may utilize  coupling  between additional waves in their components, thereby engineering the behavior of the resultant coupled modes, and enabling their coupling with free-space light. 

The origin of the new primordial metamaterial behavior can be uncovered if we follow the evolution of the electromagnetic response of a (weakly) nonlocal composite as the size of its components is gradually reduced from the macroscopic wavelength scale to the deeply subwavelength nonlocality scale. While the transition to primordial response is a universal property of any composite material, we illustrate this transition with the example of stratified, layered media comprising nonlocal components. 

\begin{figure*} 
	\centering
	\includegraphics[width=0.9\textwidth]{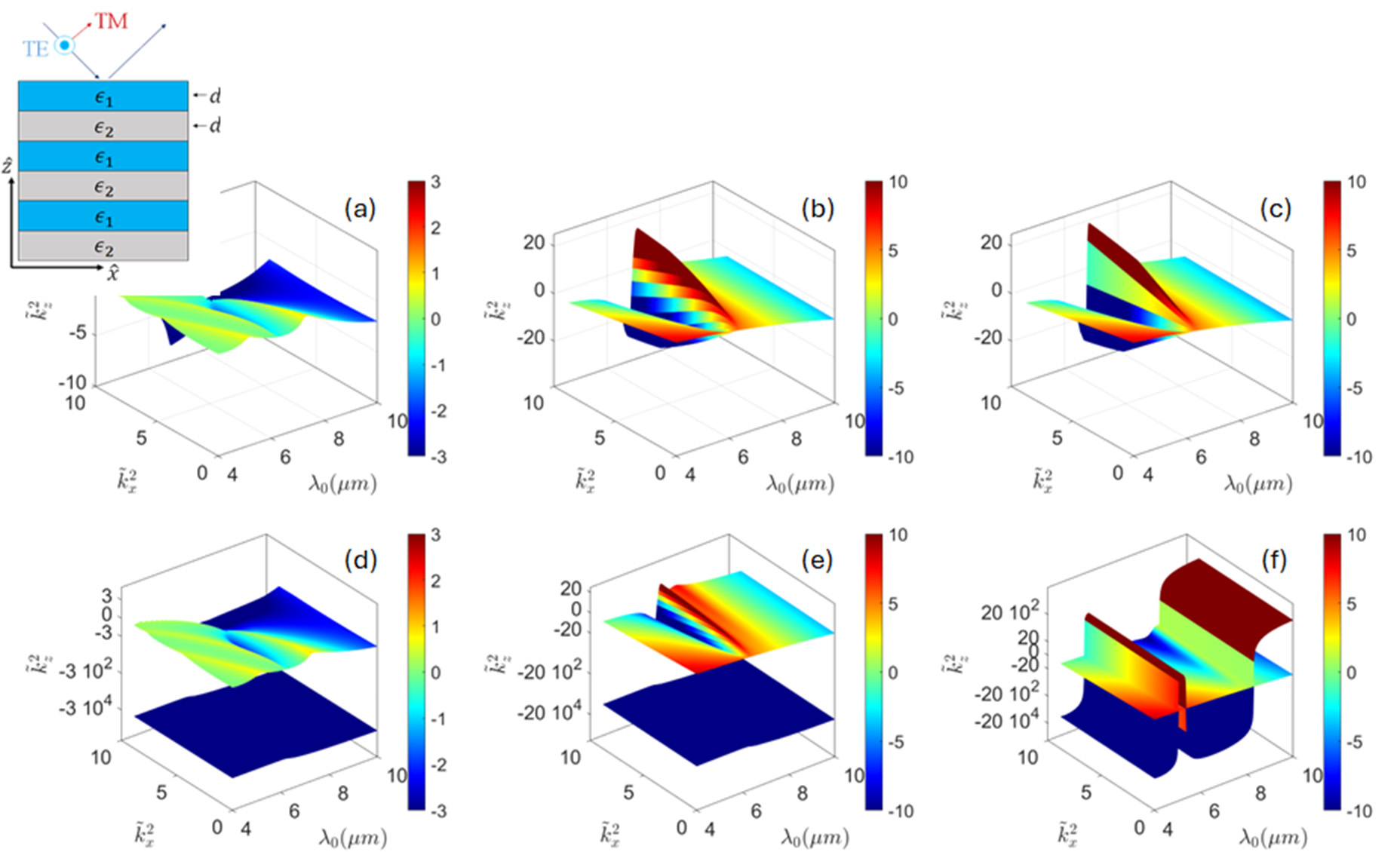} 

	\caption{\textbf{Electromagnetic modes in local and nonlocal periodic bilayer meta-materials}
		Dispersion of the modes present in periodic bi-layered materials in local (a...c) and nonlocal (d...f) periodic bi-layer metamaterials [see inset in (a)]. Components of type 1 and 2 represent plasmonic and dielectric materials; permittivity of the plasmonic components are shown in Fig.\ref{figConfig} while permittivity of the dielectric components are fixed at $\epsilon=10$ for local calculations and at $\epsilon=10+(10^{-4}+3\cdot 10^{-5}i)\tilde{k}_z^2$ for nonlocal calculations. The layer thicknesses are $d=1.5\mu m$ (a,d), $d=250 nm$ (b,e), and $d=10 nm$ (c,f); (note the nonlinear scale of the vertical axis in d...f) }
	\label{figPC} 
\end{figure*}

\begin{figure*} 
	\centering
	\includegraphics[width=1.0\textwidth]{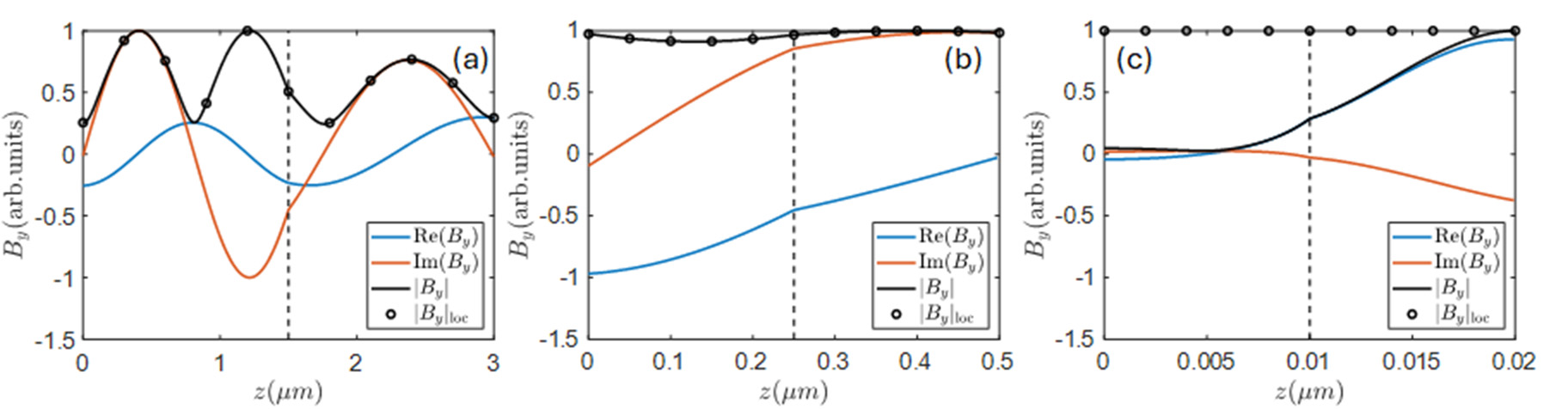} 

	\caption{\textbf{Field distributions across the composites}
		Distribution of magnetic field of the fundamental (largest $\tilde{k}_z^2$) mode across one period of composites shown in Fig.\ref{figPC} for $\lambda_0=5.05\mu m$. Solid lines correspond to nonlocal calculations; symbols represent calculations that neglect nonlocality; panels (a,b,c) correspond to layer thicknesses of $1.5\mu m$, $250 nm$, and $10 nm$, respectively.}
	\label{figFields} 
\end{figure*}

Following the transfer-matrix formalism, originally developed to analyze the dispersion of the modes in periodic arrays of electromagnetically-local layers\cite{YehTMM}, we develop a nonlocal transfer matrix method (see SI for details) and use the developed approach to analyze the propagating properties of the waves in periodic electromagnetically nonlocal arrays. Our results are summarized in Fig.\ref{figPC} that shows the dispersion of modes in bi-layer structures of fixed composition as a function of operating wavelength and in Fig.\ref{figFields} which illustrates the field distribution in the fundamental (largest $\tilde{k}_z^2$) mode across the unit cell of the composite for fixed composition and operating wavelength. 

When the layers are substantially thick, the structure operates as a one-dimensional photonic crystal, exhibiting a series of transparent propagation bands and opaque photonic band-gaps\cite{YehTMM}. Incorporating nonlocality into the model reveals an additional mode that exponentially decays into the composite. Therefore, as expected, the overall behavior of the periodic layered material is not affected by the nonlocality. Analysis of field distribution across the unit cell of the composite further confirms this conclusion, revealing the field variation on the scale of (thick, wavelength/4-scale) layers caused by the interference (Fig. \ref{figFields}a). Note that both the dispersion of the main wave and the field distribution of this wave are perfectly described by local electromagnetism.    

As the layer thickness is reduced and the individual layers become sub-wavelength, the optical response of the composite converges to the behavior of a homogeneous material whose local dielectric permittivity is well-described with effective medium theory\cite{podolskiyNonlocalAPL}. In this regime the multilayer composite behaves as a uniaxial material, with an optical axis perpendicular to the layers\cite{hoffman}. The fields remain relatively homogeneous across the composite in this regime (Fig. \ref{figFields}b). As before, incorporating nonlocality in this regime results in an additional evanescent wave and does not affect the behavior of the primary mode. 

However, when the layer size is reduced further and becomes comparable to the scale of its components' nonlocality, the electromagnetic response of the composite undergoes a qualitative change. The nonlocal theory now predicts the existence of the additional propagating mode and a significant deviation of the dispersion of the main wave from the (local) effective medium response. 

A composite operating in this {\it primordial metamaterial} regime is inherently nonlocal -- its optical response is not adequately described by local permittivity and the composite as a whole does not behave as a quasi-homogeneous effective-medium material. Notably, despite the small layer thickness, the electromagnetic fields of these primordial modes vary on the scale of the layer (Fig. \ref{figFields}c), opening the door to novel mechanisms for molding light at the nanoscale.


The presence of multiple interacting nonlocal components in the material is of vital importance for the existence of observable, propagating, primordial waves. For example, as the nonlocality of one of the layers is decreased, the behavior of the bi-layer composite converges back to the predictions of local effective medium theory. 

As mentioned above, the emergence of the primordial response, illustrated here on the example of a bi-layer periodic material, is a universal property of composite media whose components are structured on a length scale commensurate with the scale of their nonlocal response. As a result, primordial nonlocality can be realized in other coupled nonlocal environments (such as two- and three-dimensional periodic materials, nonlocal nano-cavities comprising nonlocal core and cladding layers, and interface regions).

This inherently nonlocal electromagnetism, introduced  here, is fundamentally different from previously-reported phenomena such as local hyperbolic response\cite{hoffman}, strong coupling-metamaterials\cite{Sohr_strong_coupling}, and ballistic resonances\cite{ballisticOptica,ballisticOL}, all of which are adequately described by local material parameters. Indeed, the ballistic resonance originates from the confinement of free electron motion and manifests itself at scales smaller than the nonlocality scale while strong coupling-induced phenomena\cite{Sohr_strong_coupling,jacobStrongCoupling} are observed in the long-wavelength hyperbolic regime and for large angles of incidence.
In all these previous studies, ultra-thin multilayer composites behaved as a homogeneous slab of material with some effective permittivity. In contrast, primordial composites feature fast oscillations of electromagnetic fields on the scale of their components and cannot be described by effective medium parameters. 

\subsection*{Experimental demonstration of primordial metamaterials}

Experimental realization of primordial metamaterials requires a composite with components that have (i) macroscopic (multi-atom, $\sim$ nm) nonlocality scale, (ii) high quality interfaces (with typical roughness smaller than the nonlocality scale), and (iii) uniformity of component size across the composite. These requirements can be satisfied in an epitaxially-grown semiconductor platform.

\begin{figure*} 
	\centering
	\includegraphics[width=1.0\textwidth]{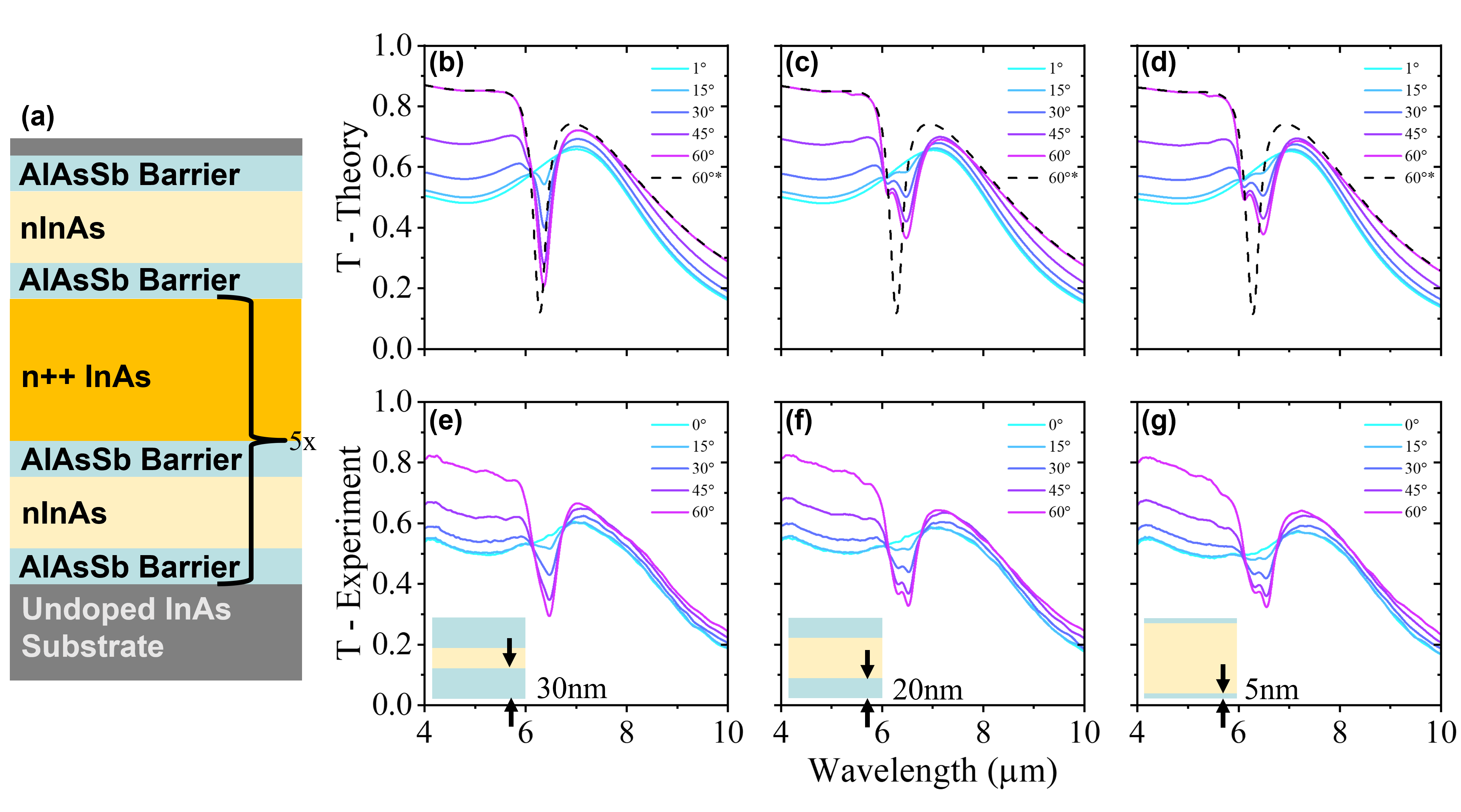} 

	\caption{\textbf{Barrier Thickness Study of Primordial HMMs}
		Layer stack of the metamaterial studies grown for this is shown in a). TM-polarized transmission spectra were simulated, (b-d), and measured, (e-g), for HMMs of varying barrier thickness and constant period. Our HMM system is comprised of the periodic semiconductor material stack AlAsSb/n-InAs/AlAsSb/n\textsuperscript{++}-InAs: (a,d) 30nm/20nm/30nm/80nm, (b,e) 20nm/40nm/20nm/80nm, and (c,f) 5nm/70nm/5nm/80nm. Insets show the variation of the barrier and well thicknesses between samples; The dashed lines in (b-d) represent $60^\circ$ transmission simulated using purely local permittivities; note the absence of the double dip in transmission in the local TMM predictions.}
	\label{xfig1} 
\end{figure*}

Semiconductor plasmonic designer metals whose nonlocality originates from their free-electron dynamics (see SI) can be epitaxially grown, with excellent control of layer thickness, free electron concentration, and interface quality, and thus present the material of choice for our study. Depending on the concentration of free electrons, such materials can exhibit transparent dielectric, reflective (plasmonic) metal, or even epsilon-near-zero response\cite{Law:12,ENZfunnel}. The primordial metamaterial platform realized in our work is schematically shown in Fig.\ref{xfig1}(a). The highly doped InAs (n\textsuperscript{++}-InAs) layers are designed to exhibit a plasmonic transition at $\lambda_p\simeq 6.3\mu m$, while their lightly-doped ``dielectric'' counterparts (n-InAs) exhibit a similar transition at approximately $17 \mu m$.  These nonlocal plasmonic layers are separated by thin undoped AlAsSb barriers, used to implement the ``hard'' interfaces that exist in our theoretical description. 
The doping of the n-InAs dielectric layer was chosen to bring its permittivity in line with the permittivity of the AlAsSb barriers at the operating wavelength $\lambda_0\sim \lambda_p$. Therefore, from the standpoint of electromagnetism, the lightly-doped InAs and the AlAsSb barrier should exhibit almost identical behavior, differing only in their nonlocal response.

Molecular beam epitaxy (MBE) was used to grow several series of metamaterials consisting of 5 periods of the structures described above. In the first set of samples, we varied the size of the AlAsSb barriers between 5 and 30 nm (keeping the total thickness of the AlAsSb/n-InAs/AlAsSb structure at 80 nm) equal to the thickness of the highly doped InAs layers, aiming to vary the strength of the nonlocal response within the composite. The angle- and wavelength-resolved transmission of this set of composites is shown in Fig. \ref{xfig1}.  The composite with the thickest AlAsSb barrier exhibits behavior that is almost identical to the previously-reported optical response of all-semiconductor-based, strongly-anisotropic (hyperbolic) metamaterials\cite{hoffman}, with a strong angle-dependent transmission feature at the wavelength where the permittivity of the highly doped layers (and the effective permittivity of the composite in the direction of its optical axis) is close to zero. However, as the thickness of the AlAsSb barrier is reduced, the optical behavior undergoes both quantitative and qualitative changes: the single angle-dependent transmission dip spectrally broadens and splits into two. Notably, such splitting in transmission (or reflection) resonances has been previously identified as a clear signature of a strongly nonlocal response in low-temperature homogeneous media\cite{kiselevJETP73,Hopfield63} and in plasmonic metamaterials\cite{podolskiyNonlocalPRL}. 

The optical response of the composites was modeled with a nonlocal transfer matrix method, where the permittivity of the doped layers (both n-InAs and n\textsuperscript{++}-InAs) were approximated as those of (nonlocal) degenerate plasmas\cite{landauPK,stockmanImperfectLens} and the AlAsSb layers were assumed to have frequency-independent permittivity with layer-thickness-dependent nonlocality (see SI). The results of our theoretical models closely match the experimental data.

\begin{figure*} 
	\centering
	\includegraphics[width=1.0\textwidth]{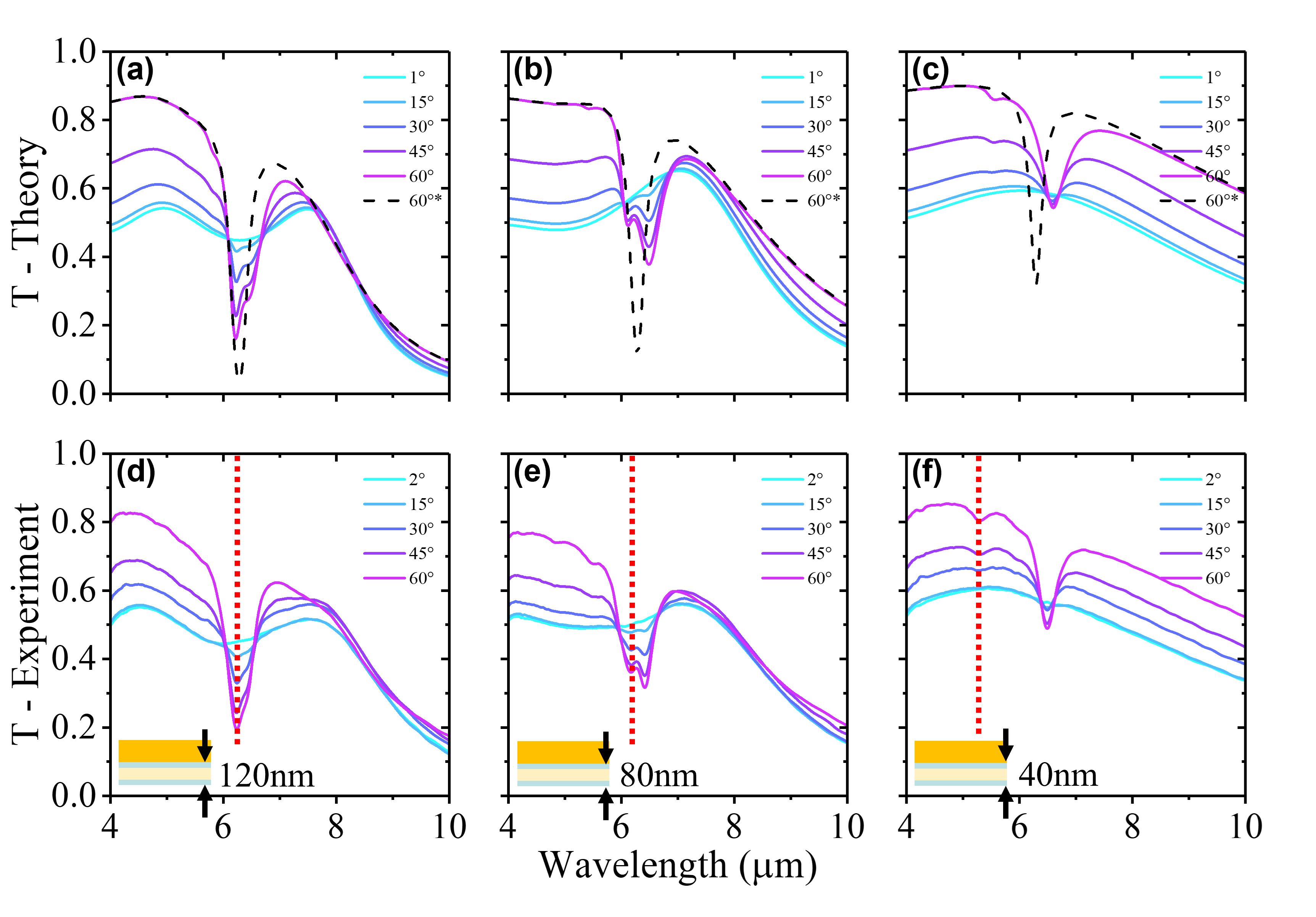} 

	\caption{\textbf{Period Thickness Study of Primordial HMMs}
		TM-polarized transmission spectra were simulated, (a-c), and measured, (d-f), for HMMs of varying period and constant barrier thickness. AlAsSb/nInAs/AlAsSb/n\textsuperscript{++}InAs: (a,d) 5nm/110nm/5nm/120nm, (b,e) 5nm/70nm/5nm/80nm, and (c,f) 5nm/30nm/5nm/40nm. Insets show the variation of the period thickness between samples; dashed lines in (a-c) represent predictions of local calculations for $60^\circ$ incidence; dashed line in (d-f) illustrates position of nonlocal spectral feature.}
	\label{xfig2} 
\end{figure*}

In the second set of experiments, we kept the thickness of the AlAsSb layers constant (thereby fixing the nonlocal response of the individual components of the metamaterial), but changed the thickness of the doped layers of the structure, varying the period of the primordial metamaterial from 80 to 240 nm. Results of these studies are illustrated in Fig.\ref{xfig2}(d-f), along with results of theoretical modeling(a-c). The effective medium descriptions of the metamaterials with varying (deep subwavelength) period should be identical. Instead, a dramatic change of the optical response of the composite is observed as the metamaterial unit cell period is decreased by a factor of three, from \SI{240}{\nm} ($\sim\lambda_0/30$) to \SI{40}{\nm} ($\sim\lambda_0/80$).  This drastic modulation of the overall electromagnetic response illustrates one of the central properties of primordial metamaterial response: even in the limit when individual layers are deeply subwavelength (here, layer thickness $\sim \lambda_0/100$), the composite does not behave as a homogeneous effective material.

Figure \ref{xfig2} illustrates another unique property of primordial metamaterials: by changing the scale of their components, it becomes possible to control nonlocality-induced optical signatures outside the previously-explored epsilon-near-zero frequency range.  This is clearly observed in the shortest period structure of Fig. \ref{xfig2}(c,f), where the signature of the nonlocal mode appears at wavelengths well below the ENZ feature at \SI{6.5}{\um}.

\subsection*{Discussion and Conclusions}

In this work we, for the first time, proposed and realized a macroscopic material that integrates multiple coupled nonlocal components and supports novel primordial waves. We have demonstrated, theoretically and experimentally, that structuring material at the deep subwavelength scale comparable with the scale of its components' nonlocal response can dramatically change the electromagnetic response of the material as a whole. The electromagnetic properties of the resulting media are strongly influenced by the primordial nonlocal interactions of the components, regardless of whether the components by themselves or the material as a whole exhibit epsilon near zero response -- a known pre-requisite to all previous realizations of nonlocal materials\cite{kiselevJETP73,AgranovichBook,wubsNonlocal}. Importantly, primordial metamaterials exhibit strong nonlocality even at room temperature and with realistic (lossy) materials while their optical properties cannot be adequately described by homogenized bulk material parameters even when the individual components are much smaller than the operating wavelength.

Primordial metamaterials provide a novel platform for controlling electromagnetic fields at the scale of individual components, affecting all aspects of light matter interactions -- from bulk properties (demonstrated in this work) to light emission, detection, and nonlinear light interactions.



%
\bibliography{Primordial_Modes} 

%
%
%
%
%
%


\section*{Acknowledgments}
\paragraph*{Funding:}
This research was supported by NSF DMREF program (award \#2118787 (VP), award \#2119157 (EN), award \#2119302 (DW\&SB))

\paragraph*{Author contributions:}
AW was responsible for growth and optical characterization of samples, JLM - for implementing nonlocal transfer matrix and theoretical simulations, all authors discussed the data and worked on the manuscript. 

\paragraph*{Competing interests:}
There are no competing interests to declare.

\paragraph*{Data and materials availability:}
All data are available in the manuscript or the supplementary materials. All data will be provided upon request to the authors.





\newpage

\onecolumngrid
\renewcommand{\thefigure}{S\arabic{figure}}
\renewcommand{\thetable}{S\arabic{table}}
\renewcommand{\theequation}{S\arabic{equation}}
\renewcommand{\thepage}{S\arabic{page}}
\setcounter{figure}{0}
\setcounter{table}{0}
\setcounter{equation}{0}
\setcounter{page}{1} 


\begin{center}
\section*{Supplementary Materials }

A. Ware, J. LaMountain$^{}$, R.C. White, S.R. Bank, \and  D. Wasserman$^{}$, V.A. Podolskiy$^{\ast}$\\ 
\small$^\ast$Corresponding author. Email: viktor\_podolskiy@uml.edu\\
\end{center}


\subsection*{Materials and Methods}

\subsubsection*{Nonlocal response of composites in real- and wavenumber- spaces}

We assume that electromagnetic radiation propagates in the $xz$ plane of the Cartesian coordinate system, that material properties change only along the $z$ direction, and that materials exhibit nonlocality primarily along the same direction. Under these circumstances, the electric displacement field in the material is related to the electric field via 
\begin{equation}
\vec{D}={\epsilon}\vec{E}-\frac{c^2}{\omega^2} \frac{\partial}{\partial z}\left(\alpha(z)\frac{\partial E_z}{\partial z}\right)\hat{z} 
\end{equation}
with $\epsilon$ and $\alpha$ describing local and nonlocal contributions to material permittivity, respectively.

Any monochromatic field propagating in the medium can then be represented as a linear combination of transverse electric (TE, $\vec{E}\|\hat{y}$) and transverse magnetic (TM, $\vec{B}\|\hat{y}$) modes, with the latter affected by the nonlocal response and being the main focus of this study. 

In this geometry it can be shown\cite{podolskiyNarimanovNonlocal}  that Maxwell equations result in the following differential equation for magnetic field in TM modes: 
\begin{equation}
\label{eqBy}
    \left(\epsilon_{}-\frac{c^2}{\omega^2}\frac{\partial}{\partial z}\alpha(z)\frac{\partial}{\partial z}\right)
    \left[\frac{\partial}{\partial z}\left\{ \frac{1}{\epsilon} \frac{\partial B_y}{\partial z}\right\} +\frac{\omega^2}{c^2}B_y\right]-k_x^2 B_y=0. 
\end{equation}

When Eq. (\ref{eqBy}) is applied to homogeneous materials, its solutions yield\cite{podolskiyNarimanovNonlocal} a set of plane waves with $E,B\propto \exp(-i\omega t+i \vec{k}\cdot\vec{r})$ whose dispersion is given by 
\begin{equation}
\label{eqNonlocalHom}
    \alpha \frac{k_z^4 c^4}{\omega^4}+(1-\alpha) \epsilon_{}\frac{k_z^2 c^2}{\omega^2}
    +\epsilon_{}\left(\frac{k_x^2 c^2}{\omega^2}-\epsilon_{}\right)=0, 
\end{equation}
and whose constitutive relationship reduces to  $D_{\beta}=\sum_\gamma\left(\epsilon \delta_{\beta\gamma} +\delta_{\beta \gamma}\alpha \frac{k_z^2 c^2}{\omega^2}\delta_{\gamma z}\right)E_{\gamma}$ with subscripts corresponding to Cartesian coordinates and $\delta_{\beta\gamma}$ being the Kroenecker delta symbol. 

The application of Eq. (\ref{eqBy}) to an interface between two homogeneous materials yields a set of boundary conditions\cite{podolskiyNarimanovNonlocal} requiring continuity of $E_x$, $B_y$, $E_z$, and $\alpha \frac{\partial E_z}{\partial z}$ across the interface. Note that the two former conditions represent the conventional boundary conditions of local electromagnetism, while the two latter relationships represent the additional boundary conditions that are required to fully solve for coupling between primary and additional waves at the interface between two nonlocal media. Continuity of $E_z$ then becomes redundant at the local/nonlocal interface.

\subsubsection*{Nonlocal transfer matrix formalism}

The transfer matrix formalism is a powerful linear algebra technique for calculating the propagation of light through an arbitrary array of planar layers. In our implementation, the (column vector) of the fields at a given point inside a layer $l$ is given by the product of the layer-specific structure matrix $\hat{N_l}$, a diagonal phase matrix $\hat{F_l}(z)$, and a vector of mode-specific amplitudes $\vec{a}_l$. Explicitly, 

\begin{eqnarray}
\hat{N}_l=
\begin{bmatrix}
    1 & 1 & 1 & 1 \\
    \frac{\epsilon_l}{\tilde{k}_{z_{l,1}}} & \frac{\epsilon_l}{\tilde{k}_{z_{l,2}}} &
    -\frac{\epsilon_l}{\tilde{k}_{z_{l,1}}} & -\frac{\epsilon_l}{\tilde{k}_{z_{l,2}}} \\
    \alpha_l({\tilde{k}^2_{z_{l,1}}-\epsilon_l}) &
    \alpha_l(\tilde{k}^2_{z_{l,2}}-\epsilon_l) &
    \alpha_l(\tilde{k}^2_{z_{l,1}}-\epsilon_l) &
    \alpha_l(\tilde{k}^2_{z_{l,2}}-\epsilon_l) \\

    \frac{\tilde{k}^2_{z_{l,1}}-\epsilon_l}{\tilde{k}_{z_{l,1}}} &
    \frac{\tilde{k}^2_{z_{l,2}}-\epsilon_l}{\tilde{k}_{z_{l,2}}} &
    -\frac{\tilde{k}^2_{z_{l,1}}-\epsilon_l}{\tilde{k}_{z_{l,1}}} &
    -\frac{\tilde{k}^2_{z_{l,2}}-\epsilon_l}{\tilde{k}_{z_{l,2}}} 
\end{bmatrix}
\\
\hat{F_l}(z)=
\begin{bmatrix}
    \exp(i k_{z_{l,1}}z) &0 &0 &0 \\
    0& \exp(i k_{z_{l,2}}z) &0 &0 \\
    0& 0& \exp(-i k_{z_{l,1}}z) &0 \\
    0& 0& 0& \exp(-i k_{z_{l,2}}z),    
\end{bmatrix}
\end{eqnarray}
with $\epsilon_l,\alpha_l$ representing material parameters of the layer $l$, $k_{z_l,m}$ describing the propagation constant of the mode $m=1,2$ in the layer $l$ [see Eq.(\ref{eqNonlocalHom})], and $\tilde{k}_{z_l,m}=c k_{z_l,m}/\omega$. The rows of the field matrix $\hat{N}_l$ represent distributions of $E_x$, $B_y$,  $\alpha \frac{\partial E_z}{\partial z}$, and $E_z$ fields, respectively while matrix $\hat{F}_l$ describes the phase of individual modes.

The boundary conditions at the interface between the layers $l$ and $l+1$, located at the point $z_0$ can then be written as 
\begin{equation}
    \hat{N_l}\hat{F_l}(z_0)\vec{a_l}=\hat{N}_{l+1}\hat{F}_{l+1}(z_0)\vec{a}_{l+1}, 
\end{equation}
that allows one to relate the amplitudes of the modes in layer $l+1$ to the amplitudes of the modes in layer $l$ via the transfer matrix $\hat{T}_l$: $\vec{a}_{l+1}=\hat{T}_l \vec{a}_{l}$, with $\hat{T}_l=(\hat{N}_{l+1}\hat{F}_{l+1}(z_0))^{-1}\hat{N_l}\hat{F_l}(z_0)$. 

Similar to the (local) transfer matrix method (TMM), originally introduced in Ref.\cite{YehTMM}, the nonlocal TMM can be used to calculate transmission, and reflection of light by layered optical stacks with arbitrary layer configuration\cite{podolskiyNarimanovNonlocal}. 

The nonlocal TMM can also be used to calculate dispersion and field profiles of the modes propagating in periodically stratified materials. In the case of bi-layer periodic materials with thicknesses $d_1$ and $d_2$, the Bloch-periodic condition reduces to: 
\begin{equation}
    \left(\hat{F}_1(d_1)(\hat{N}_1)^{-1}\hat{N}_2 \hat{F}_2(d_2) (\hat{N}_2)^{-1}\hat{N}_1-
    \exp[i q(d_1+d_2)]\hat{I}\right)\vec{a}_1=0, 
\end{equation}
with $q$ being the wavenumber of the mode, $\vec{a}_1$ describing the magnitudes of the fields in the first layer, and $\hat{I}$ being the identity matrix. Dispersions of the modes propagating in periodically stratified nonlocal bilayer media are shown in Figs.\ref{figPC},\ref{figFields} of the main manuscript. Sample Matlab code is available in \cite{githubPrimordial}.

\subsubsection*{Nonlocal material parameters of epitaxially grown layers}

The permittivity of the doped semiconductors is approximated by the permittivity of a degenerate collisionless plasma\cite{landauPK}, corrected for the contribution from the background permittivity of the semiconductor. Explicitly, 
\begin{equation}
    \epsilon_{{\rm pl}_{\alpha \beta}}(\omega, \vec{k})=\epsilon_\infty\left[
    \delta_{\alpha\beta}\left(1-\frac{\omega_{\rm pl}^2}{\omega(\omega+i\gamma)}\right)
    -\frac{3}{5}\delta_{\alpha z}\delta_{\beta z}\frac{\omega_{\rm pl}^2(k_z v_F)^2}{\omega^3(\omega+i\gamma_{nl})}
    \right] 
\end{equation}
For doped InAs materials, we use background permittivity $\epsilon_\infty=12.3$, $\omega_p$ corresponding to plasma wavelengths of $6.3\mu m$ and $17.2\mu m$, respectively, and calculate Fermi velocity based on Fermi energy and electron effective mass $v_F=\sqrt{2E_f/m_*(E_f)}$. The latter parameters are estimated using fully quantum mechanical calculations, previously described in Refs.\cite{ballisticOptica,ballisticOL} that derive optical response of the highly doped semiconductor layers based on discrete quantum well transitions and that incorporate mass non-parabolicity. Importantly, these calculations demonstrate that the layer thicknesses in our materials are large enough for the Drude approach to adequately describe the response of the highly doped layers. Inelastic damping parameters are set at $\gamma=8 THz,\gamma_{nl}=28.3THz$.  

In contrast to previous works on semiconductor designer-metal-based metamaterials\cite{hoffman,ballisticOL,ballisticOptica}, the optical response of the thin AlAsSb barriers used in this work is strongly affected by the highly inhomogeneous charge distribution, resulting from charge tunneling between the two neighboring doped layers. 

Full description of this behavior will require first-principles calculations to understand the microscopic charge distribution, as well as quantum mechanical calculations assessing the spatial dependence of local and nonlocal permittivities across the layer. Notably, given the high doping used in this work, significant non-parabolicity of effective mass, and significant expected band bending, such calculations will likely require fitting parameters to match the results of first-principles calculations to experimental data. 

As an alternative, effective nonlocality of the  barrier can be calculated analytically, by minimizing the energy functional
\begin{equation}
\Phi[E]=\frac{1}{2}\int_{z_i}^{z_f}dz\left\{\epsilon(z)E(z)^2-
\alpha(z)\left(\frac{\partial E}{\partial z}\right)^2\right\}-
D_0(z) \int_{z_i}^{z_f}dz E(z), 
\end{equation}
with respect to effective parameters $\epsilon_{\rm eff}$ and $\alpha_{\rm eff}$, with $D_0(z)=\left\langle\epsilon_{\rm eff} E(z)+\alpha_{\rm eff}\frac{\partial^2 E}{\partial z^2}\right\rangle$ and $\langle\cdots\rangle$ representing spatial average across the layer. 

While this approach requires knowledge of boundary conditions $E(z_i)$ and $E(z_f)$, an adequate estimate can be obtained by introducing a trial function $E(z)\approx E_t(z)=D_0/\epsilon_{b}+A \cos\left(\sqrt{\frac{\epsilon_b}{\alpha_{\rm eff}}}z\right)$, with $A=\left(E_0-\frac{D_0}{\epsilon_b}\right)/\cos\left(\sqrt{\frac{\epsilon_b}{\alpha_{\rm eff}}}\frac{d}{2}\right)$, with $\epsilon_b$ being local permittivity of the barrier material, and $d$ being barrier thickness. 

The above minimization procedure yields:
\begin{eqnarray}
    \alpha_{\rm eff}\simeq \langle \alpha \rangle+ \frac{2\delta \alpha}
    {d\sqrt{\langle \epsilon\rangle\langle\alpha\rangle}\sin\left(\sqrt{\frac{\langle\epsilon\rangle}{\langle \alpha\rangle}}\frac{d}{2}\right)}
    \label{eqAlpEff}
\end{eqnarray}
with $\delta\alpha$ being variation of nonlocality within the layer. 

Given that the exact distributions of material permittivity -- that is determined by the exact distribution of charge density across the barrier layer --  cannot be derived from the experiment, and given the extreme sensitivity of the effective nonlocality on these distributions [see Eq.(\ref{eqAlpEff})], here we utilize a single fitting parameter to describe the (thickness-dependent) nonlocality of the barrier layers. Specifically, we use $\epsilon_b=10$ and $\alpha_{\rm eff}=(2.75+1.5i)\times10^{-5}$ for 5-nm barriers; nonlocality of 20- and 30-nm-thick barriers is reduced by 10\% and 75\%, respectively.

\subsubsection*{Metamaterial growth}
All metamaterial samples presented in this work were grown via molecular beam epitaxy using a Varian Gen-II system on 1/4 2" unintentionally doped InAs substrates. All samples were subjected to a load chamber bake at \SI{500}{\celsius} for 2.5 hours after initial loading then transferred to a buffer chamber and baked individually at \SI{300}{\celsius} for 1 hour. Prior to growth, each sample was transferred to the substrate manipulator of the growth chamber and baked at \SI{300}{\celsius} for 15 min. After this initial bake, the substrate manipulator was rotated to the growth position and a final bake at \SI{510}{\celsius} was performed under Arsenic flux to remove the surface oxide from the substrate. Prior to the growth of the epitaxial stacks shown in this work, a \SI{200}{\nm} InAs buffer layer was grown at \SI{475}{\celsius} to provide an epitaxial surface on which initiate our metamaterial growth and thus reduce defect densities in the following layers\cite{sheldon_buffers}. The remainder of the growth was maintained at \SI{475}{\celsius}. The lightly n-type doped InAs layers (n-InAs) were doped using a GaTe source with intended doping concentration of $N_D = 2.09 \times 10\textsuperscript{18}~cm\textsuperscript{-3}$, corresponding to a plasma wavelength $\lambda_{p} = 17.2~\mu m$. The degenerately-doped InAs layers (n\textsuperscript{++}-InAs) were doped with a Silicon source. The approximate doping concentrations extracted from the transmission spectra were $N_D = 3.00 \times 10\textsuperscript{19}~cm\textsuperscript{-3}$, corresponding to a plasma wavelength $\lambda_{p} = 6.3~\mu m$. Following the growth of the metamaterial stack, the samples were capped with \SI{10}{\nm} of InAs to prevent oxidation of the Aluminum containing layers.

\subsubsection*{Metamaterial characterization}
\begin{figure} 
	\centering
	\includegraphics[width=1.0\textwidth]{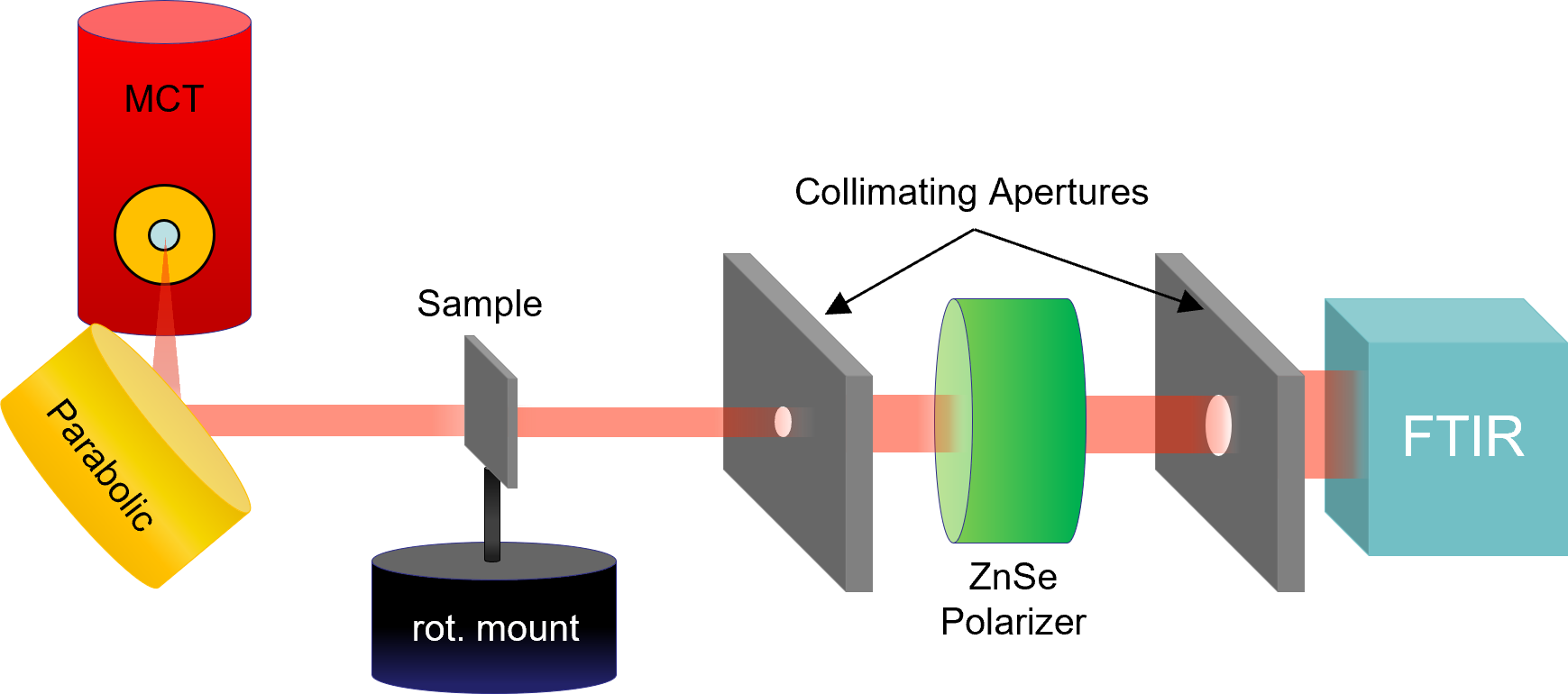} 

	\caption{\textbf{Experimental Transmission Setup}
		}
	\label{fig:S1} 
\end{figure}

All samples grown for this work were characterized using polarization and angle-dependent Fourier Transform Infrared (FTIR) transmission spectroscopy. The experimental system used is shown in Fig. \ref{fig:S1}. Broadband MIR light from an internal globar source was passed through the internal interferometer of the FTIR. After exiting the FTIR, this light was passed through a series of collimating apertures as well as a ZnSe wire-grid polarizer. Samples were mounted vertically on a rotating stage, aligned to rotate about the point where light is incident on the sample, allowing variation of the incidence angle. Light transmitted through the samples was collected using a $2"$ focal length parabolic gold mirror and focused onto a Mercury Cadmium Telluride detector (MCT). All transmission measurements shown in this work are normalized to transmission through air using the same experimental system, with the sample removed.

\begin{figure} 
	\centering
	\includegraphics[width=1.0\textwidth]{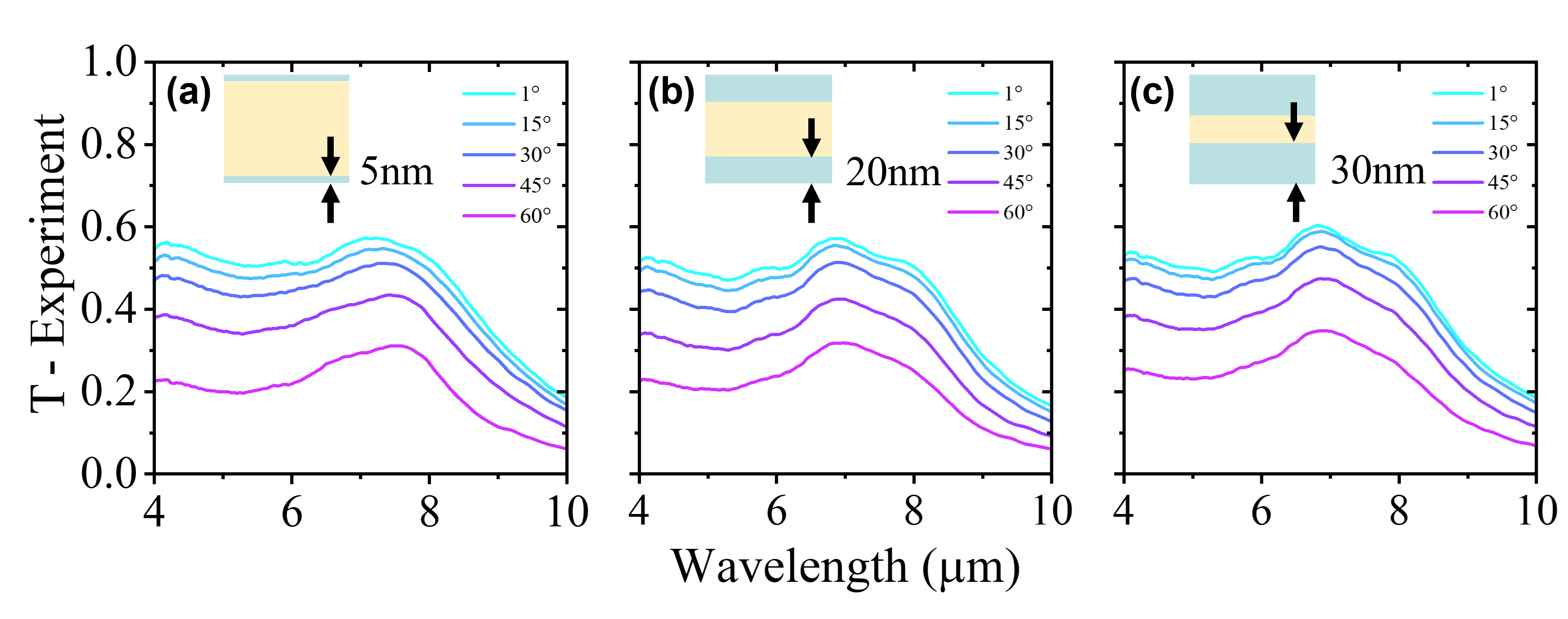} 

	\caption{\textbf{TE Polarized Barrier Study}
		TE-polarized transmission spectra for HMMs of varying barrier thickness and constant period. AlAsSb/nInAs/AlAsSb/n\textsuperscript{++}InAs: (a,d) 5nm/70nm/5nm/80nm, (b,e) 20nm/40nm/20nm/80nm, and (c,f) 30nm/20nm/30nm/80nm. Insets show the variation of the barrier thickness between samples.}
	\label{fig:S2} 
\end{figure}

Fig.\ref{fig:S2} shows the experimental transmission spectra for TE-polarized light of the three samples used in the barrier-thickness study. It is seen that, in contrast to the results shown in Fig. \ref{xfig1}, the transmission of TE-polarized light is almost independent of the barrier thickness, illustrating that the in-plane material response (which should not affected by the nonlocality of the metamaterial layers) is almost identical across all samples.  









\end{document}